\documentclass{emulateapj}
\usepackage{apjfonts}
\usepackage{color}

\definecolor{darkbrown}{RGB}{139,69,19}

\lefthead{HAN ET AL.} 
\righthead{OGLE-2015-BLG-0051/KMT-2015-BLG-0048L\MakeLowercase{b}}

\begin{document}

\title{OGLE-2015-BLG-0051/KMT-2015-BLG-0048L\MakeLowercase{b}: a Giant Planet Orbiting a Low-mass Bulge Star Discovered by High-cadence Microlensing Surveys}

\author{
C.~Han$^{1}$, A.~Udalski$^{2,9}$, A.~Gould$^{3,4,10}$, V.~Bozza$^{5,6}$,\\
and \\
Y.~K.~Jung$^{1,9}$, M.~D.~Albrow$^{7}$, S.-L.~Kim$^{4}$, C.-U.~Lee$^{4}$, S.-M.~Cha$^{4,8}$,
D.-J.~Kim$^{4}$, Y.~Lee$^{4,8}$, B.-G.~Park$^{4}$, I.-G.~Shin$^{1,9}$    \\ (The KMTNet Collaboration),\\
M.~K.~Szyma{\'n}ski$^{2}$, I.~Soszy{\'n}ski$^{2}$, J.~Skowron$^{2}$, P.~Mr{\'o}z$^{2}$, 
R.~Poleski$^{2,3}$, P.~Pietrukowicz$^{2}$, S.~Koz{\l}owski$^{2}$, K.~Ulaczyk$^{2}$, 
{\L}.~Wyrzykowski$^{2}$, M.~Pawlak$^{2}$\\
(The OGLE Collaboration),\\
}

\affil{$^{1}$ Department of Physics, Chungbuk National University, Cheongju 361-763, Republic of Korea}
\affil{$^{2}$ Warsaw University Observatory, Al. Ujazdowskie 4, 00-478 Warszawa, Poland}
\affil{$^{3}$ Department of Astronomy, Ohio State University, 140 W. 18th Ave., Columbus, OH 43210, USA}
\affil{$^{4}$ Korea Astronomy and Space Science Institute, Daejon 305-348, Republic of Korea}
\affil{$^{5}$ Dipartimento di Fisica "E. R. Caianiello", Universit'a di Salerno, Via Giovanni Paolo II, 84084 Fisciano (SA), Italy}
\affil{$^{6}$ Istituto Nazionale di Fisica Nucleare, Sezione di Napoli, Via Cintia, 80126 Napoli, Italy}
\affil{$^{7}$ University of Canterbury, Department of Physics and Astronomy, Private Bag 4800, Christchurch 8020, New Zealand}
\affil{$^{8}$ School of Space Research, Kyung Hee University, Yongin 446-701, Republic of Korea}
\affil{$^{9}$ Harvard-Smithsonian Center for Astrophysics, 60 Garden St., Cambridge, MA, 02138, USA}
\footnotetext[9]{The OGLE Collaboration}
\footnotetext[10]{The KMTNet Collaboration}

\begin{abstract}
We report the discovery of an extrasolar planet detected from the combined data
of a microlensing event OGLE-2015-BLG-0051/KMT-2015-BLG-0048 acquired by
two microlensing surveys. Despite that the short planetary signal occurred in
the very early Bulge season during which the lensing event could be seen for just
about an hour, the signal was continuously and densely covered.  From the 
Bayesian analysis using models of the mass function, matter and velocity 
distributions combined with the information of the angular Einstein radius,
it is found that the host of the planet is located in the Galactic bulge. 
The planet has a mass $0.72_{-0.07}^{+0.65}\ M_{\rm J}$ and it is orbiting a low-mass 
M-dwarf host with a projected separation $d_\perp=0.73 \pm 0.08$ AU.  
The discovery of the planet demonstrates the capability of the current 
high-cadence microlensing lensing surveys in detecting and characterizing planets.
\end{abstract}


\section{INTRODUCTION}

Since the first discovery by \cite{wolszczan92} followed by \cite{mayor95},  
many exo-planets have been discovered. With the {\it Kepler} mission, the number 
of known planets explosively increased and now exceeds $\sim$3000
according to the Extrasolar Planets Encyclopaedia\footnote{\tt http://exoplanet.edu}.
Most of them were discovered by either the transit, e.g.\ \citet{Tenenbaum14},
or radial-velocity methods, e.g.\ \citet{Pepe11}.

Planets have also been discovered using the microlensing method.  Due to the fact 
that these planetary systems are detected through their gravitational fields rather 
than their radiation, this method makes it possible to detect planets around faint 
stars and even dark objects. Furthermore, microlensing is sensitive to planets in 
wide orbits beyond the snow line, which separates regions of rocky planet formation 
from regions of icy planet formation, while other major planet detection techniques 
are sensitive to close-in planets. Being able to detect planets that are difficult 
to be detected by other techniques, the method is important for the comprehensive 
understanding of planet formation \citep{gaudi12}.

The number of known microlensing planets at the time of writing this paper is 46,
which is relatively small compared to the number of planets detected by other major 
methods.  There are two main reasons for the small number of microlensing planets.  
The first reason is the rarity of microlensing events.  The optical depth to microlensing, 
which represents the average probability of a star to be gravitationally lensed at a given 
moment, toward the Galactic bulge field is of order $10^{-6}$ \citep{sumi03, sumi06}.  
Then, observation cadences of survey experiments were limited because they should monitor 
a large area of sky in order to maximize the number of observing stars.  The second reason 
is the short duration of planetary microlensing signals.  A planetary companion to a stellar 
lens exhibits its presence through a short-term perturbation to the smooth and symmetric 
lensing light curve induced by the host star \citep{mao91, gould92b}.  It was difficult 
to cover such short planetary signals by early-generation lensing surveys that had 
$\sim 1/2$ -- 1 day observation cadences.  To detect short planetary signals, 
earlier lensing experiments adopted a strategy where lensing events were detected by 
wide-field surveys and events detected by surveys were intensively monitored using multiple 
narrow-field telescopes \citep{albrow98}.  However, only a small fraction of ongoing events, 
which exceeds several hundreds during an observing season, could be observed by approximately 
a dozen follow-up telescopes.  As a result, the detection efficiency of microlensing planets 
under the survey/follow-up mode observation had been low.

However, past few years have witnessed great changes in microlensing surveys.  With the 
start of the fourth phase survey experiment, the Optical Gravitational Lensing Experiment 
\citep[OGLE:][]{udalski15} group significantly increased the observation cadence by 
broadening the field of view of their camera from  $0.4\ {\rm deg}^2$ into $1.4\ {\rm deg}^2$.  
The Microlensing Observation in Astrophysics \citep[MOA:][]{bond01} group also plans to 
upgrade their camera to widen the current $2.2\ {\rm deg}^2$ field of view into 
$4\ {\rm deg}^2$ (T.~Sumi, private communication).  There were additions of instruments 
to microlensing surveys.  The Wise team \citep{shvartzvald14} joined microlensing surveys 
in 2011 by using its 1.0m telescope.  The Korea Microlensing Telescope Network (KMTNet) 
survey, that is composed of 3 globally distributed telescopes equipped with large-formation 
cameras, started microlensing observation in 2015 season.  With the continuous and dense 
coverage of lensing events achieved by the instrumental upgrade of existing survey groups 
and the addition of new surveys, microlensing planet search is entering a new phase where 
planets can be detected by survey-mode observations alone.

In this paper, we report the discovery of a giant planet from the joint data acquired 
by the OGLE and KMTNet survey experiments.  The short-lasted planetary signal occurred 
in the very early Bulge season during which the event could be seen for just about an hour. 
Nevertheless, the signal was densely and continuously covered by the 2 surveys experiments, 
enabling to detect and characterize the planetary system.

The paper is organized as follows. In Section 2, we describe the observation of the planetary
microlensing event by  survey experiments and acquired data. In Section 3, we give a description 
about the modeling procedure conducted to analyze the observed lensing light curve. We provide the 
estimated physical parameters of the discovered planetary system in Section 4. 
Finally, we summarize results and make a brief discussion about the result in Section 5.

\section{OBSERVATION AND DATA}

The planet was discovered from the observation of a microlensing event OGLE-2015-BLG-0051
that occurred on a star located toward the Galactic bulge field. The equatorial coordinates 
of the lensed star (source) are (RA,DEC) = 
($17^{\rm h}58^{\rm m}39^{\rm s}\hskip-2pt.01$,$-28\arcdeg01\arcmin54\arcsec\hskip-2pt.1$),
that correspond to the Galactic coordinates ($l$,$b$) = ($2.24\arcdeg$,$-2.00\arcdeg$).

The event was discovered by the OGLE Early Warning System \citep[EWS:][]{udalski15} on 
February 13, 2015 from observations using the 1.3m Warsaw telescope at the Las Campanas 
Observatory in Chile.  On March 2, 2015, it was noticed that the event experienced an 
anomaly and an alert was issued to the microlensing community for follow-up observations.  
However, the alert was issued when the anomaly was almost finished and thus the major part 
of the anomaly could not be covered by follow-up observations.

The event was analyzed in real time with its progress.  From the modeling conducted 
by one of us (CH) during the anomaly, it was pointed out that the anomaly was possibly 
of planetary origin, although other binary interpretations could not be completely 
excluded.  Continued modeling conducted after the anomaly by CH and other modelers 
(VB and MDA) reached a consistent result that the anomaly was produced by a planetary 
companion to the lens.

Although the event could not be observed by follow-up observations, it was 
densely observed by 
the KMTNet lensing survey that is designed to monitor a large area of the Galactic bulge field 
with high cadences by using large-format cameras equipped on multiple telescopes. 
The KMTNet survey
started its test observation in February, 2015, which 
matches the occurrence time of the event.  The event was dubbed as KMT-2015-BLG-0048 
in the KMT event list.\footnote{{\tt http://astroph.chungbuk.ac.kr/$\sim$kmtnet}}  The survey 
uses three identical telescopes that are located at Cerro Tololo Interamerican Observatory 
in Chile (KMT CTIO), South African Astronomical Observatory in South Africa (KMT SAAO), 
and Siding Spring Observatory in Australia (KMT SSO). At the time of the event, KMT SSO 
was not online and the event was observed by two telescopes, KMT CTIO and KMT SAAO. Each 
telescope has a 1.6m aperture and is equipped with a mosaic camera composed of four 
9K$\times$9K CCDs. Each CCD has a pixel size of 10 microns corresponding to 0.4 arcsec/pixel 
and thus the camera has a 4 ${\rm deg}^2$ field of view \citep{kim16}. 
For the 
major fields, the observation cadence of the survey is $\sim 10$ min. This cadence is high 
enough to detect signals produced by Earth-mass planets considering that the perturbation 
time of the signal is $\sim 3$ hr.

In our analysis, we use combined data acquired by the OGLE and KMTNet surveys. 
The OGLE data are composed of  1167 $I$-band images.  The KMTNet data consist of 786 
$I$-band and 54 $V$-band images obtained from KMT CTIO observations and 1117 $I$-band 
images acquired from KMT SAAO observations.  The main use of the KMT CTIO $V$-band data 
is to constrain the source star but they are not used for the light curve analysis because 
(1) the number of data is small and (2) the photometry is relatively poor due to extinction.
There exist data taken 
by the MOA group but we do not use them in our analysis not only because the perturbation 
region covered by the data overlaps with that covered by the combined OGLE+KMTNet data but 
also because the photometry quality is relatively poor.

\begin{figure}[t]
\epsscale{1.15}
\plotone{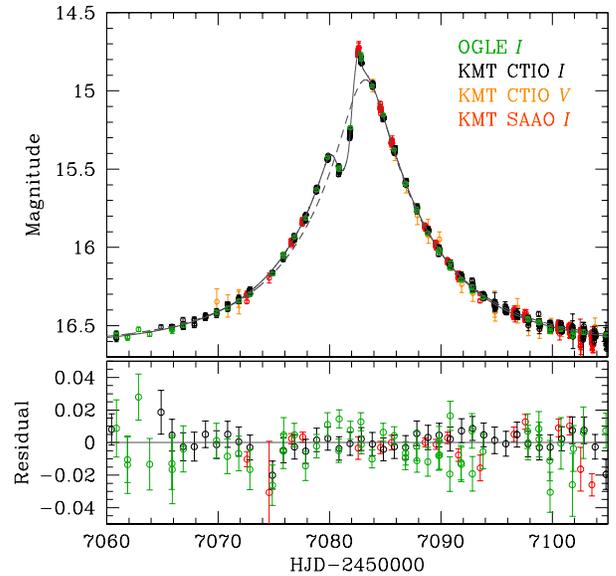}
\caption{\label{fig:one}
Light curve of the microlensing event OGLE-2015-BLG-0051/KMT-2015-BLG-0048.
The solid curve superposed on the data is the model light curve 
based on the planetary model, while the dashed curve is based on the 
point-source point-lens (PSPL) model. 
Residuals of the KMT data sets are daily binned.
}
\end{figure}

Photometry of the images are conducted using the customized pipelines of the individual 
groups.  Both pipelines are based on the Difference Imaging Analysis method 
\citep{alard98, wozniak00, albrow09}.  Since data are taken by different telescopes and 
processed by different photometry codes, we renormalize error bars of the individual data 
sets by 
\begin{equation} 
\sigma ' = k ( {\sigma_0^2} + {\sigma_{\rm min}^2} ) ^{1/2} ,
\label{eq1}
\end{equation}
where $\sigma_0$ is the error estimated from the pipeline, $\sigma_{\rm min}$ is a factor 
used to make the cumulative distribution function of $\chi^2$ as a function of lensing 
magnification linear, and $k$ is a scaling factor to make $\chi^2$ per degree of freedom 
become unity.  Photometric precision improves as the source star is magnified and 
the factor $\sigma_{\rm min}$ is needed to make the scatter of data points be consistent 
with the error bars of the source brightness. The scaling factor $k$ is needed to ensure 
that each data set is fairly weighted according to its error bars.  We note that the 
error-bar normalization parameters vary as a lensing model varies.  We iterate the normalization 
process and the final parameters are set when the model is stable. In Table~\ref{table:one}, 
we present the estimated normalization parameters of the individual data sets. Although 
the event was observed in the early season, we find no systematic trend in the photometry 
caused by airmass trends.

\begin{deluxetable}{lcc}
\tablecaption{Error-bar normalization parameters\label{table:one}}
\tablewidth{0pt}
\tablehead{
\multicolumn{1}{c}{Data set}     &
\multicolumn{1}{c}{$k$}          &
\multicolumn{1}{c}{$\sigma_{\rm min}$ (mag)} 
}
\startdata
OGLE      &   1.749   & 0.001 \\
KMT CTIO  &   1.386   & 0.005 \\
KMT SAAO  &   1.162   & 0.002 
\enddata
\end{deluxetable}

In Figure~\ref{fig:one}, we present the light curve of OGLE-2015-BLG-0051/KMT-2015-BLG-0048. 
Compared to the continuous and symmetric light curve of a single-mass event, the light curve 
exhibits a short-term perturbation during $7080.0\lesssim{\rm HJD}-2450000\lesssim7082.5$. 
The perturbation shows a feature that is composed of a depression centered at 
${\rm HJD}-2450000\sim7081.5$ and brief bumps at both edges of the depression.  Such dips, 
usually surrounded by two bumps, are a generic feature of lensing systems with small mass 
ratios $q\ll1$ and normalized planet-star separations $s<1$, i.e., planets inside the 
Einstein ring, which represents the source image caused by the exact alignment of the source, 
lens, and observer.  When a source is lensed by the host of a planet, the host star's gravity 
generates two images, one inside and the other outside the Einstein ring.  The former, being 
a saddle point on the time delay surface, is easily suppressed if the planet lies in or near 
the path, thereby causing relative demagnification, and hence a dip in the light curve 
\citep{gaudi12}.

Besides the main feature of the anomaly, there appears to exist a weak anomaly at 
${\rm HJD}\sim 2457075$, where 4 data points show a $\sim 0.02$ mag level deviation.
We consider it as a fluctuation in data because 
(1) the deviation is consistent with $3\sigma$ level of photometry,
(2) the region is sparsely observed, and
(3) a 2-body model cannot explain both of this weak and the main anomaly features.

\begin{figure}[t]
\epsscale{1.15}
\plotone{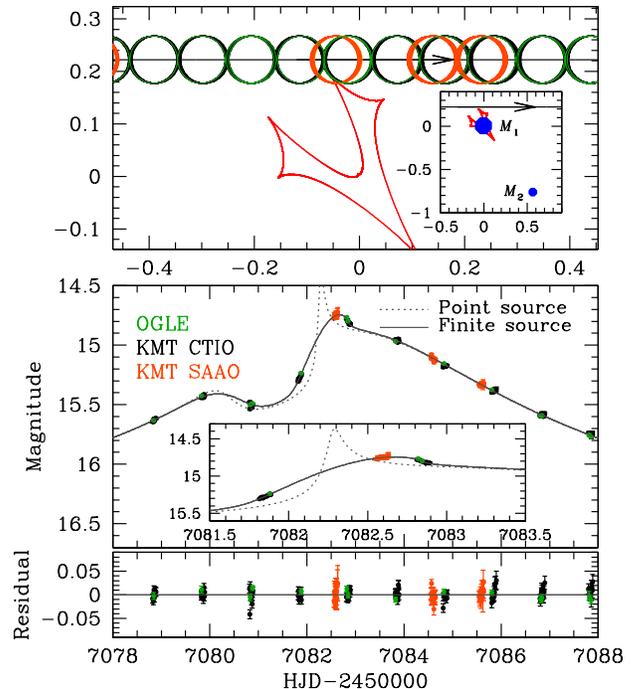}
\caption{\label{fig:two}
Lens system geometry. The upper panel shows the source
trajectory (straight line with an arrow) with respect to the lens
components (marked by $M_1$ and $M_2$) and caustics (closed concave curve) 
and the lower panel shows the light variation with the
progress of the source position. Lengths are scaled to the Einstein radius
and the source trajectory is aligned so that the progress of the source
matches the light curve shown in the lower panel. The inset in the upper
panel shows the wide view and the major panel shows the enlarged view
around the caustic. The empty circles on the source trajectory represent
the source positions at the times of observation and the size indicates the
source size. The dotted curve in the lower panel is the light curve
expected for a point source.
The inset in the middle panel shows the zoom of the light curve affected 
by finite-source effects.
}
\end{figure}

The major structure of the anomaly feature was well covered by the survey data despite the 
short time window toward the field.  See the zoom of the light curve around the planetary 
perturbation presented in Figure~\ref{fig:two}.  During the time of the perturbation when 
the Bulge field could be seen only for approximately an hour, the OGLE survey obtained 2 
images per night and the KMTNet survey obtained up to 20 images per night using its 
two telescopes. Since the OGLE and KMT CTIO telescopes are located at the sites with similar 
longitudes, the coverage of the perturbation by the two telescopes are similar. Although 
the KMT SAAO data missed the depression part of the perturbation due to poor weather 
conditions, they cover the second bump thanks to  $\sim 6.1$-hour longitude difference from 
the Chilean telescopes.

\section{ANALYSIS}

Keeping in mind that the anomaly pattern is likely to be produced by a binary lens with 
a low mass ratio, we conduct binary-lens modeling.  For the description of a binary-lensing 
light curve, one needs 7 principal parameters for the lensing system and 2 flux parameters 
for each observatory.  The first 3 of the principal parameters describe the source approach 
with respect the lens, including the time of the closest source approach to a reference 
position of the lens, $t_0$, the lens-source separation at that moment, $u_0$ (impact 
parameter), and the time for the source to cross the angular Einstein radius $\theta_{\rm E}$ 
of the lens, $t_{\rm E}$ (Einstein time scale).  For the reference position of the lens, we 
use the barycenter of the binary lens. The other 3 principal parameters describe the binary 
lens including the projected separation $s$ and the mass ratio $q$ between the binary components, 
and the angle between the source trajectory and the binary axis, $\alpha$. We note that the 
impact parameter $u_0$ and the binary separation $s$ are normalized to $\theta_{\rm E}$.  
The other parameter defined as the ratio of the angular source radius to the Einstein radius, 
$\rho = \theta_\ast / \theta_{\rm E}$, is needed to describe light curve deviations 
affected by finite-source effects.  For the graphical presentation of the binary lensing 
parameters, see Figure 6 of \citet{jung15}.  The flux parameters $f_s$ and $f_b$ 
represent the fluxes from the source and blend, respectively.

For some lensing events, observed data exhibit subtle residuals from the best-fit model 
based on the principal lensing parameters due to higher-order effects.  The known causes of 
such deviations include the parallax effect \citep{gould92a} and the lens orbital effect 
\citep{albrow00, an02, jung13}.  The parallax effect is caused by the positional change 
of the observer due to the orbital motion of the Earth around the Sun.  On the other hand, 
the lens-orbital effect is caused by the positional change of the lens due to its orbital 
motion.  Such effects are important for long time-scale events where the duration of the 
event comprises a significant fraction of the orbital period of either the Earth or the 
lens.  For OGLE-2015-BLG-0051/KMT-2015-BLG-0048 with an Einstein time scale 
$t_{\rm E}\sim 11$ days, we find that these higher-order effects are negligible.

We proceed light-curve modeling in several steps.  In the first step, we conduct a 
preliminary grid search for solutions in the parameter space of ($s$, $q$, $\alpha$), 
for which lensing light curves vary sensitively to the change of the parameters. In 
this process, other parameters, for which light curve varies smoothly to the change of 
the parameters, are searched for by using a downhill approach.  For the downhill 
$\chi^2$ minimization, we use the Markov Chain Monte Carlo (MCMC) method.  
The ranges of the $s$ and $q$ parameters inspected by the grid search are
$-1.0 < \log s \leq 1.0$ and $-4.0 < \log q \leq 1.0$, respectively, and they
are devided into $70\times 70$ grids.  
The range of the source trajectory angle is $0 < \alpha \leq 2\pi$ and it is 
divided into 15 grids.  We note that $\alpha$ is allowed to vary from each starting point
while $s$ and $q$ are fixed during the model search. 
In the second 
step, we investigate possible local solutions in the parameter space in order to check the 
existence of degenerate solutions where different combinations of the lensing parameters 
result in similar light curves.  In this process, we refine local solutions by allowing 
all parameters, including the grid parameters $s$, $q$ and $\alpha$ in the preliminary 
search, to vary.  Finally, we search for the global solution by comparing $\chi^2$ values 
of the identified local solutions.

\begin{figure}[t]
\epsscale{1.15}
\plotone{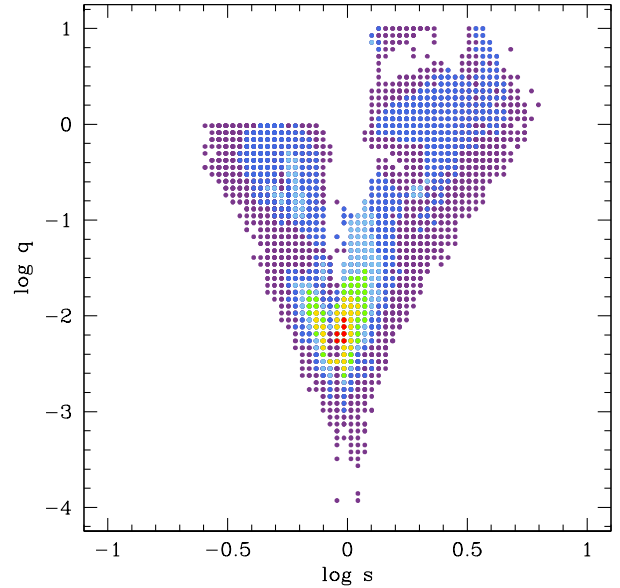}
\caption{\label{fig:three}
Map of $\Delta\chi^2$ in the $(s,q)$ parameter space obtained from the preliminary 
grid search for lensing solutions.  Color coding represents $n\sigma$ (red),  $2n\sigma$ 
(yellow),  $3n\sigma$ (green),  $4n\sigma$ (cyan),  $5n\sigma$ (blue), and  $6n\sigma$ 
(purple), where $n=20$.  We note that the preliminary search is done before error-bar 
normalization and thus $\sigma$ levels are different from the map presented in 
Fig.~\ref{fig:four}.  
}
\end{figure}

\begin{deluxetable}{lc}
\tablecaption{Lensing parameters\label{table:two}}
\tablewidth{0pt}
\tablehead{
\multicolumn{1}{c}{Parameters}     &
\multicolumn{1}{c}{Values} 
}
\startdata
$\chi^2$                   &    3057.1 \\     
$t_0 \ ({\rm HJD})$        & 2457083.081 $\pm$ 0.003 \\     
$u_0$                      &       0.224 $\pm$ 0.002 \\     
$t_{\rm E} \ ({\rm days})$ &      10.81  $\pm$ 0.07  \\     
$s$                        &       0.963 $\pm$ 0.002 \\     
$q \ (10^{-3})$            &       7.43  $\pm$ 0.13  \\     
$\alpha \ ({\rm rad})$     &       5.358 $\pm$ 0.002 \\     
$\rho \ (10^{-3})$         &      45.3   $\pm$ 0.6   \\
$f_b/f_s$                  &      0.01 $\pm$ 0.01
\enddata
\end{deluxetable}

Lensing magnifications are affected by finite-source effects when the source is located 
close to or over caustics, which represent the positions on the source plane where the 
lensing magnification of a point source becomes infinite. Caustics of binary lenses form 
a single or multiple closed curves where each curve is composed of concave curves that 
meet at cusps.  For the computation of lensing magnifications affected by finite-source 
effects, we use the ray-shooting method.  In this method, rays are uniformly shot from 
the lens plane, bent by the lens equation, and then collected on the source plane to make 
a ray map.  The lens equation of a binary lens is expressed as 
\begin{equation}
\zeta=z-
{\epsilon_1 \over \bar{z}-\bar{z}_{{\rm L},1}} - 
{\epsilon_2 \over \bar{z}-\bar{z}_{{\rm L},2}},
\label{eq2}
\end{equation}
where $\epsilon_i$ is the mass fraction of each lens component, $\zeta$, $z$, and 
$z_{\rm L,i}$ denote the positions of the source, image, and lens expressed in complex 
notation in units of the angular Einstein radius, respectively, and $\bar{z}$ represents 
the complex conjugate of $z$ \citep{witt90}.  Once a ray map is constructed, a 
finite-source magnification for a given source position is computed as the number density 
ratio of rays arrived on the source surface to the ray density on the image plane.  In the 
initial grid search, we apply the map-making method \citep{dong06}, where a single map for 
a combination of the binary parameters $s$ and $q$ is used to produce many light curves 
resulting from different source trajectories.  In the step to refine local solutions, the 
map-making method cannot be used because the parameters $s$ and $q$ are allowed to vary.  
In order to accelerate computation, we first apply semi-analytic hexadecapole approximations 
\citep{gould08, pejcha09} except when the source is on the caustic.  We also minimize the 
number of rays by shooting rays that will arrive at regions around the caustic.  
Finally, we use customized codes developed for parallel computing, where multiple 
CPUs simultaneously compute model magnifications for the individual data points instead 
of computing the magnification of each data point one by one.

In computing finite-source magnifications, we consider the surface-brightness variation of 
the source star. For this, we model the surface-brightness profile as 
\begin{equation}
S_\lambda \propto 1-\Gamma_\lambda\left( 1-{3\over 2}\cos \phi\right),
\label{eq3}
\end{equation}  
where $\Gamma_\lambda$ is the linear limb-darkening coefficient and $\phi$ is the angle 
between the line of sight toward the source and the normal to the source surface.  We adopt 
$\Gamma_I=0.46$ from \citet{claret00} based on the de-reddened color and brightness of the 
source.  See section 4 for details about the procedure to determine the color and magnitude.

\begin{figure}[t]
\epsscale{1.15}
\plotone{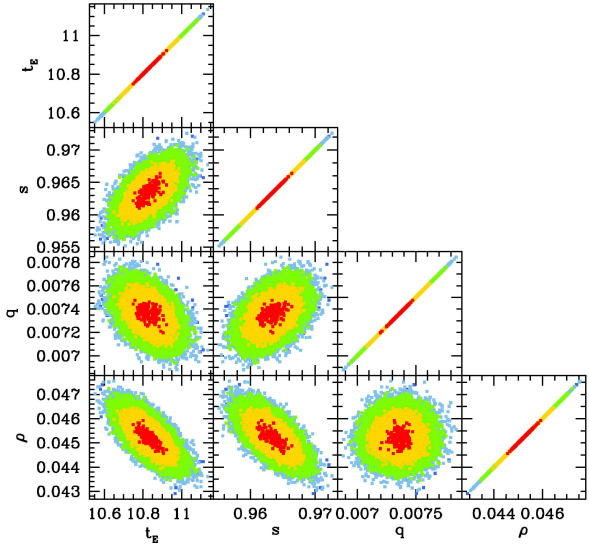}
\caption{\label{fig:four}
Distributions of the lensing parameters. 
The color coding represents points on the MCMC chains within 
$1\sigma$ (red), $2\sigma$ (yellow), $3\sigma$ (green), $4\sigma$ (cyan), and $5\sigma$ (blue) of the 
best-fit value. 
}
\end{figure}

From the search for a solution, we find a unique solution with a companion/primary 
mass ratio corresponding to a planetary case.  We find no degenerate solution with 
a $\chi^2$ value that is comparable to the best-fit solution.  This can be seen 
in Figure~\ref{fig:three}, where we present the $\Delta\chi^2$ map in the $(s,q)$ 
parameter space obtained from the preliminary grid search.   By refining the solution, 
it is estimated that the planet/host mass ratio is $q = (6.80\pm 0.18)\times10^{-3}$ and 
the projected planet-host separation is $s=0.954\pm 0.004$. In Table~\ref{table:two},
we present the best-fit lensing parameters.  In Figure~\ref{fig:four}, we also 
present the distributions of the lensing parameters on the MCMC chain in order to 
show the covariances between the lensing parameters.\footnote{
Those who want to reanalyze the event can download the MCMC chain and the light 
curve data at\\
 http://astroph.chungbuk.ac.kr/$\sim$cheongho/OB150051/.} 
The MCMC run is stopped by visually inspecting the posterior distributions in the 
parameter space.
The uncertainty of each parameter is estimated from the scatter of the MCMC chain.

We note that the shape of the light curve resembles that of MOA-2007-BLG-192 
\citep{Bennett08}.  For MOA-2007-BLG-192, the observational coverage of the planetary 
deviation is sparse and incomplete and thus there exist multiple possible solutions. 
On the other hand, the coverage of the deviation of OGLE-2015-BLG-0051 is dense and 
complete thanks to high-cadence observation from multiple distributed locations, 
leading to unambiguous characterization of the planetary system.

In the upper panel of Figure~\ref{fig:two}, we present the geometry of the lens system 
corresponding to the best-fit solution. Due to the resonance of the projected separation 
to $\theta_{\rm E}$, i.e.\ $s\sim1$, the lens system forms a single caustic around the host 
of the planet.  The source passed the backside of the arrowhead-shaped caustic.  The depression 
in the light curve occurred when the source was in the demagnification valley between the two 
protrudent cusps that caused the brief bumps on both sides of the depression. The source crossed 
the tip of one of the cusps during which the light curve shows a clear finite-source signature 
from which we accurately measure the normalized source radius $\rho$.  Considering that 
normalized source radii for typical lensing events produced by low-mass stars located roughly 
halfway between the source and observer are $\rho \sim 10^{-3}$ for events occurred on 
main-sequence stars and $\sim$10$^{-2}$ even for events involved with giant source stars, 
the measured value of $\rho = (45.2\pm 0.8)\times10^{-3}$ is unusually large.  For a 
given size of a bulge star, this suggests that the Einstein radius is very small.

\section{PHYSICAL PARAMETERS}

Since finite-source effects are clearly detected, it is possible to determine the angular 
Einstein radius from the relation 
\begin{equation}
\theta_{\rm E} = {\theta_\ast \over \rho}.
\label{eq4}
\end{equation}
The normalized source radius $\rho$ is measured from the analysis of the light curve around 
the planetary perturbation.  The angular radius of the source star, $\theta_\ast$, is 
estimated from the source type that is determined based on the de-reddened color $(V-I)_0$ 
and brightness $I_0$.

\begin{figure}[t]
\epsscale{1.15}
\plotone{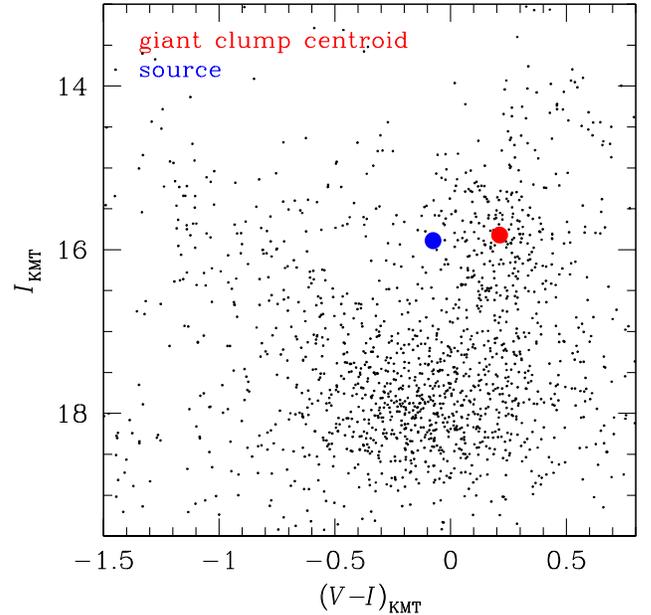}
\caption{\label{fig:five}
Source position in the instrumental color-magnitude diagram 
of nearby stars with respect to the centroid of the giant clump.
The diagram is constructed based on stars in the KMT subfield ($140''\times 140''$ area)
including the source star.
}
\end{figure}

In order to determine the calibrated color and brightness of the source star, we use the 
method of \citet{yoo04}.  Following the method, we first locate the source star in the 
instrumental (uncalibrated) color-magnitude diagram of neighboring stars in the same field.
We then calibrate the color and brightness from the offsets between the positions of the 
source star and the centroid of the giant clump (GC), for which its de-reddened color 
$(V-I)_{0, \rm GC}$ and brightness $I_{0, \rm GC}$ are known \citep{nataf13}.  By adopting 
$(V-I)_{0, \rm GC}=1.06$ \citep{bensby11} and $I_{0, \rm GC}=14.4$ accounting for a variation 
with Galactic longitude \citep{nataf13}, we find that $(V-I, I)_0 = (0.77\pm 0.05, 14.5\pm 0.03)$.  
This indicates that the source is a G-type giant star.  We then convert $V-I$ into $V-K$ using 
the color-color relation of \cite{bessel88} and finally determine $\theta_\ast$ using the 
color-angular radius relation of \cite{kervella04}.  We find that the angular source radius 
is $\theta_\ast=4.40\pm0.38\ {\rm{\mu}as}$.  We note that the two principal sources of 
uncertainty in estimating $\theta_*$ are the uncertainty of the dereddened color, 
$\sigma (V-I)_0 \sim 0.05$ mag, and the mag uncertainty in the determined position of GC, 
$\sim 0.1$ mag.  These two sources combined yield a $\sim 7\%$ uncertainty in the estimated 
$\theta_*$.  On the other hand, the uncertainty in the color-size relation is small compared 
to the principal sources of error \citep{Gould2014}.  The angular Einstein radius estimated 
from $\theta_*$ is then 
\begin{equation}
\theta_{\rm E}=0.093\pm0.008\ {\rm{mas}} .
\label{eq5}
\end{equation}
Combined with the Einstein time scale measured from the light curve modeling, the relative 
lens-source proper motion is determined as 
\begin{equation}
\mu={\theta_{\rm E} \over t_{\rm E}}=3.15\pm0.28{\ }{\rm{mas}{\ }yr^{-1}} .
\label{eq6}
\end{equation}
As expected from the large $\rho$, the estimated Einstein radius is significantly smaller than 
$\sim0.5$ mas for typical events produced by low-mass stars.

\begin{deluxetable}{lc}
\tablecaption{Physical lens parameters\label{table:three}}
\tablewidth{0pt}
\tablehead{
\multicolumn{1}{c}{Parameter}          &
\multicolumn{1}{c}{Value} 
}
\startdata
Mass of the host star      & $0.10_{-0.01}^{+0.09}\ M_\odot$   \\
Mass of the planet         & $0.72_{-0.07}^{+0.65}\ M_{\rm J}$  \\
Distance                   & $8.2\pm 0.9$ kpc          \\
Projected separation       & $0.73\pm 0.08$ AU         
\enddata
\end{deluxetable}

For the unique determination of the mass $M$ and distance $D_{\rm L}$ to the lens, it is 
required to measure both the lens parallax $\pi_{\rm E}$ and the Einstein radius 
$\theta_{\rm E}$ \citep{gould92a} that are related to the physical parameters by
\begin{equation}
M={\theta_{\rm E} \over \kappa \pi_{\rm E}}\qquad
D_{\rm L}={ {\rm AU}\over \pi_{\rm E}\theta_{\rm E}+\pi_{\rm S}},
\label{eq7}
\end{equation}
where $\kappa=4G/(c^2{\rm AU})\simeq 8.1~{\rm mas}/M_\odot$ and $\pi_{\rm S}={\rm AU}/D_{\rm S}$ 
is the parallax of the source star located at a distance $D_{\rm S}$.  
For OGLE-2015-BLG-0051/KMT-2015-BLG-0048, the lens parallax cannot not measured due to the 
short time scale of the event and thus the physical parameters cannot be uniquely determined.

Although unique determinations are difficult, one can statistically constrain the physical 
lens parameters based on the measured Einstein radius $\theta_{\rm E}$ and the relative 
lens-source proper motion $\mu$ combined with a Galactic model.  For this, we conduct a 
Bayesian analysis by using models of the mass function, matter and velocity distributions.
The Galactic model is based on \citet{han95}.  In this model, the matter distribution is 
based on a double-exponential disk and a triaxial bulge. The disk velocity distribution is 
assumed to be Gaussian about the rotation velocity and the bulge velocity distribution is a 
triaxial Gaussian with components deduced from the flattening via the tensor virial theorem.  
The mass function is based on the \citet{gould00} model, that includes stars, 
 brown dwarfs and stellar remnants of
whote dwarfs, neutron stars, and black holes.
Based on the the 
Galactic model, we produce a large number ($6\times 10^6$) of artificial Galactic microlensing 
events and compute the relative probability of the individual events.   From the lensing parameter 
distribution of artificial events, we then estimate the range of the lens mass and distance 
corresponding to the measured $t_{\rm E}$ and $\theta_{\rm E}$.

\begin{figure}[t]
\epsscale{1.15}
\plotone{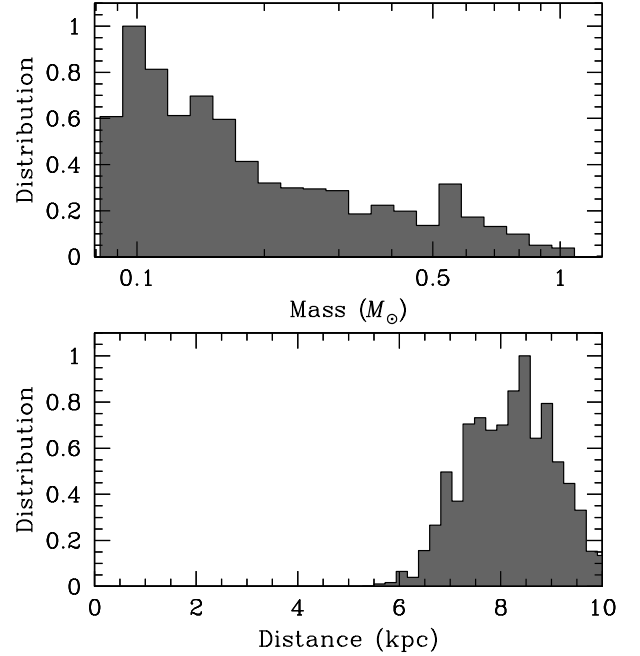}
\caption{\label{fig:six}
Posterior distributions of the lens mass and distance obtained from Bayesian analysis.
}
\end{figure}

Figure~\ref{fig:six} shows the posterior distributions of the lens mass 
and distance obtained from Bayesian analysis. 
The mass and distance to the 
lens estimated by the Bayesian analysis are
\begin{equation}
M=0.10_{-0.01}^{+0.09}\ M_\odot,\qquad   
D_{\rm L}=8.2\pm0.9 \ {\rm kpc},
\label{eq8}
\end{equation}
respectively.  
We note that the upper and lower error bars of the mass are different due to the asymmetric 
distribution of the mass distribution.
The estimated mass of the host corresponds to a low-mass star.
However, considering that the likelihood of the lens mass peaks just above the lower limit of the assumed 
mass function, one cannot exclude the possibility that the host is a brown dwarf. 
We note that the mean value of the estimated distance to the lens approximately 
coincides with the adopted distance to the Galactic center $d_{\rm GC}\sim 8\ {\rm kpc}$, implying 
that the source star is likely to be located behind the Galactic center and the lens is close 
to the Galactic center.  This result comes from the fact that (1) the bulge self-lensing 
probability is higher for source stars behind the Galactic center due to the longer line of 
sight and (2) the lens density profile peaks at the Galactic center.  
We find that the blended light provides little constraint on the mass and 
distance distributions because the flux from the host star of the planet 
is negligible compared to the flux from the source star.

The measurement of $q$ from the lens model then directly yields an estimate 
of the planet mass
\begin{equation}
M_{\rm p}=qM=
0.72_{-0.07}^{+0.65}\ M_{\rm J},
\label{eq9}
\end{equation}
while the measurement of $s$ yields the projected lens-host separation
\begin{equation}
d_\perp = s D_{\rm L} \theta_{\rm E} = 0.73 \pm 0.08 \ {\rm AU}.
\label{eq10}
\end{equation}
The snow line distance of the host star is 
$a_{\rm sl}\sim 0.5\ {\rm AU}$ according to the relation  
$a_{\rm sl}=2.7\ {\rm AU} (M/M_\odot)$ \citep{kennedy08}.
Therefore, the planet is located beyond the snow line of the host star.

\section{SUMMARY AND DISCUSSION}

We reported the discovery of an extrasolar planet that was detected from the combined data 
of a microlensing event OGLE-2015-BLG-0051/KMT-2015-BLG-0048 acquired by the OGLE and 
KMTNet surveys.  Continuous and dense coverage of the short planetary signal by the survey 
data collected despite the short time window in the early bulge season enabled unambiguous 
detection and characterization of the planetary system.  We find that the planet has a mass 
about twice that of the Jupiter and it is orbiting a low-mass host star located in the 
Galactic bulge.  The discovery of the planet well demonstrates the capability of the current 
lensing surveys with enhanced observation cadence achieved by the instrumental upgrade of 
existing surveys and the addition of new surveys.

Cool M dwarfs far outnumber sun-like stars and thus understanding the process of planet 
formation around them is important. Furthermore, small masses and low luminosities of M 
dwarfs provide leverage on conditions of planet formation, enabling to check the validity 
of existing formation theories and refine survived theories, e.g.\ \citep{ida05, boss06}.
With improved survey capability, future microlensing planet sample will include planets not 
only in greatly increased number but also in wide spectrum of hosts and planets, helping us 
to have better and comprehensive understanding about the formation and evolution of planets.

\acknowledgments
Work by C.~Han was supported by Creative Research Initiative Program (2009-0081561) of 
National Research Foundation of Korea.  
OGLE Team thanks Profs.\ M.\ Kubiak and G.\ Pietrzy{\'n}ski, former
members of the OGLE team, for their contribution to the collection of
the OGLE photometric data over the past years.
The OGLE project has received funding from the National Science Centre,
Poland, grant MAESTRO 2014/14/A/ST9/00121 to AU.
Work by AG was supported by JPL grant 1500811.
We acknowledge the high-speed internet service (KREONET)
provided by Korea Institute of Science and Technology Information (KISTI).
The KMTNet telescopes are operated by the Korea Astronomy and Space Science Institute (KASI).

\end{document}